\documentclass[aps,prl,twocolumn,superscriptaddress,showpacs]{revtex4}
\usepackage{graphicx}
\usepackage{rotating,dcolumn}
\begin{document}

\title{Spin liquid correlations in Nd-langasite anisotropic Kagom\'e antiferromagnet}

\author{J. Robert}
\affiliation{Laboratoire Louis N\'{e}el, CNRS, B.P. 166, 38 042
Grenoble Cedex 9, France}
\author{V. Simonet}
\affiliation{Laboratoire Louis N\'{e}el, CNRS, B.P. 166, 38 042
Grenoble Cedex 9, France}
\author{B. Canals}
\affiliation{Laboratoire Louis N\'{e}el, CNRS, B.P. 166, 38 042
Grenoble Cedex 9, France}
\author{R. Ballou}
\affiliation{Laboratoire Louis N\'{e}el, CNRS, B.P. 166, 38 042
Grenoble Cedex 9, France}
\author{P. Bordet}
\affiliation{Laboratoire de Cristallographie, CNRS, B.P. 166, 38
042 Grenoble Cedex 9, France}
\author{P. Lejay}
\affiliation{Centre de Recherches des Tr\`es Basses
Temp\'eratures, CNRS, B.P. 166, 38 042 Grenoble Cedex 9, France}
\author{A. Stunault}
\affiliation{Institut Laue-Langevin, BP 156, 38042 Grenoble Cedex,
France.}

\date{\today}

\begin{abstract}

Dynamical magnetic correlations in the geometrically frustrated
Nd$_3$Ga$_5$SiO$_{14}$ compound were probed by inelastic neutron
scattering on a single crystal. A scattering signal with a ring
shape distribution in reciprocal space and unprecedented
dispersive features was discovered. Comparison with calculated
static magnetic scattering from models of correlated spins
suggests that the observed phase is a spin liquid inherent to an
antiferromagnetic kagom\'e-like lattice of anisotropic Nd moments.

\end{abstract}


\maketitle Geometrically frustrated magnets are at the forefront
of research because of their diverse exotic ground states but also
because the concept of frustration plays a major role in fields
beyond magnetism \cite{ramirez}. One prototype is the kagom\'e
antiferromagnet, a two dimensional lattice of corner sharing
triangles of antiferromagnetically interacting spins, which
attracts strong attention as a promising candidate for possessing
a disordered ground state. At the quantum limit of spins 1/2,
there are converging arguments towards the stabilization of a type
II spin liquid \cite{diep} consisting in a disordered ground state
breaking neither spin nor lattice symmetry, and realizing a
quantum spin liquid state. Its search traces back to the
suggestion of a resonating valence bond state in the triangular
lattice Heisenberg model \cite{anderson}. At the classical limit
of large spins the basic signature of the geometric frustration is
the infinite degeneracy of the ground state. A disorder phase with
short range magnetic correlations, transposable to the
superposition of ground state spin configurations, can be
expected, with an intrinsic dynamical nature associated with
zero-point motions, localized or extended
\cite{kano,huse,chalker,reimers,moessner98}. Subtle differences
exist according to the spin degrees of freedom. With planar XY or
isotropic Heisenberg spins, because of the continuous extensive
entropy, thermal fluctuations play a crucial role in selecting a
subspace within the ground state manifold \cite{chalker} through
the {\it order out of disorder} mechanism \cite{villain}. Whether
non linear effects may even push forward this selection to a
unique ground state is still an open question. With Ising spins, a
cooperative paramagnet is predicted at all temperatures with a
discrete extensive ground state degeneracy and exponentially
decaying spin-spin correlations \cite{kano,barry}.

A disorder phase is experimentally easily destabilized by any kind
of small perturbation \cite{ramirez}, like second order
interactions, structural imperfections or off-stoichiometry.
Accordingly, most of the few materializations of kagom\'e
antiferromagnets available to date do order at low enough
temperatures, though often into unconventional magnetic states
(non collinear, non coplanar, multi-propagating or incommensurate
orderings, disorder free spin glassiness, ...) that are in some
cases sustaining large magnetic fluctuations. Of interest is that
these compounds (for instance the jarosites \cite{jarosite}, the
Cr kagom\'e bilayers \cite{bilayers}, the volborthite \cite{hiroi}
and the kagom\'e-staircase compounds \cite{rogado}) are best
described by Heisenberg spin models and that up to now, there was
no example of a kagom\'e antiferromagnet with a strong
magnetocrystalline anisotropy calling for an Ising or a XY
description. Another point was the absence of large enough single
crystals for three-axis inelastic neutron scattering in all cases
of kagom\'e antiferromagnets in which a spin liquid state was
suspected. This unique technique allows one to probe the dynamical
magnetic correlations both in space and time and was earlier
successfully used in evidencing signatures of spin liquid phases
in compounds containing a 3D pyrochlore net of magnetic moments :
the itinerant-electron Y$_{0.97}$Sc$_{0.03}$Mn$_2$ \cite{ballou}
and the insulating ZnCr$_2$O$_4$ \cite{lee}.

Nd-langasite, Nd$_3$Ga$_5$SiO$_{14}$, crystallizes in the trigonal
space group P321 and belongs to a family of materials that
attracted strong interest for showing better piezoelectric
properties than quartz or lithium niobate and tantalate
\cite{iwataki}. We recently discovered that some members of this
family could be also of interest as geometrically frustrated
magnets \cite{bordet}. The Nd$^{3+}$ magnetic ions in
Nd$_3$Ga$_5$SiO$_{14}$ are organized in corner sharing triangles
in well separated planes perpendicular to the {\it c}-axis. Within
these, the Nd$^{3+}$ form a distorted kagom\'e lattice, which is
topologically equivalent to the ideal kagom\'e when only the
shortest atom bridging interactions are considered. We emphasize
that this Nd net is fully occupied and that large size single
crystals can be grown.

Powders of Nd-langasite were prepared by solid state reactions of
stoichiometric amounts of high purity oxides at 1420$^{\circ}$C in
air whereas single crystals were grown from sintered powders by
the floating zone method using a mirror furnace from "Cyberstar"
company. Single-crystal magnetization measurements confirmed that
a geometric frustration does materialize in this compound and
revealed a strong magnetocrystalline anisotropy associated with
crystal field effects on the ground multiplet of the 4f$^3$
Nd$^{3+}$ ions (total angular momentum J = 9/2) \cite{bordet}. A
quantitative analysis of the magnetic susceptibility $\chi$ at
high temperatures T yielded a Curie-Weiss temperature $\theta$ =
-52~K accounting for antiferromagnetic interactions between the Nd
moments and quadrupolar crystal field parameters indicating that
these moments are coplanar rotators in the kagom\'e planes
\cite{bordet}. A change in the magnetocrystalline anisotropy
occurs around 33~K due to the higher order terms of the crystal
electric field, the {\it c}-axis, perpendicular to the kagom\'e
planes, becoming the magnetization axis at low temperature
\cite{bordet}. No phase transition towards a long-range magnetic
order was detected down to 1.6~K, a temperature much smaller than
the exchange coupling energy scale given by $\theta$, neither in
these magnetization measurements nor by elastic neutron scattering
on powders, clearly suggesting that a spin liquid phase could be
stabilized in the compound \cite{bordet}.

Inelastic neutron scattering measurements were performed at the
high flux reactor of the Institut Laue Langevin to probe the low
temperature dynamical magnetic correlation. Experiments with
powders on the IN5 time of flight spectrometer informed
essentially about the low levels of the energy spectrum of the
Nd$^{3+}$ ions as split by the crystal electric field. An
additional much weaker signal was also detected around 1 meV
\cite{robert}.

In this letter, we report on the results from an inelastic neutron
scattering experiment on a large single crystal, 40 mm long and 5
mm in diameter, performed on the cold neutron three-axis
spectrometer IN14 to probe the weak magnetic signal around 1 meV
and its distribution in the reciprocal space. The single crystal
was mounted in a cryostat with the [100] and [010] axes in the
horizontal scattering plane. Its temperature was maintained at 2
K. The neutron wavelength was selected from a vertically curved
graphite monochromator PG(002). We used a fixed final energy of
4.66 meV with energy resolution around 165 $\mu$eV. In the
following, the wave-vectors $Q$ are expressed via their
coordinates ($\xi$,$\zeta$,0) in reciprocal lattice units (r.l.u.)
or via their modulus
$|Q|/|a^*|=\sqrt{(\xi+1/2\zeta)^2+3/4\zeta^2}$ with $|a^*|$ =
0.899 \AA$^{-1}$.

\begin{figure}
\begin{center}
\includegraphics[scale=0.7]{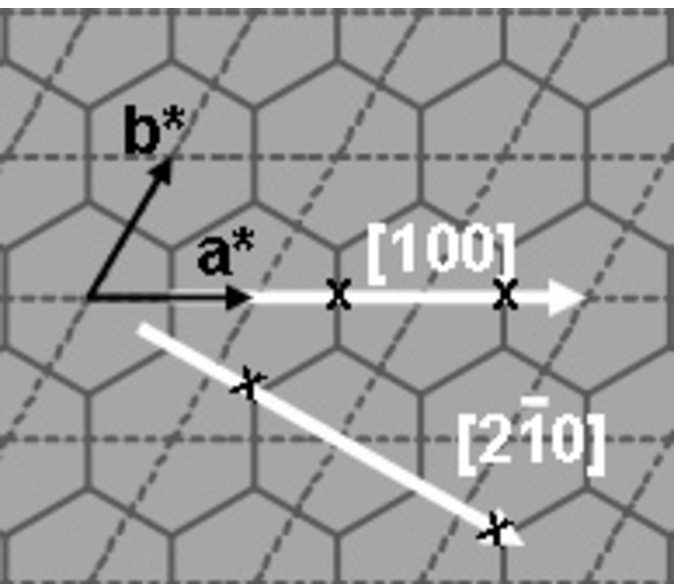}
\end{center}
\begin{center}
\includegraphics[scale=1]{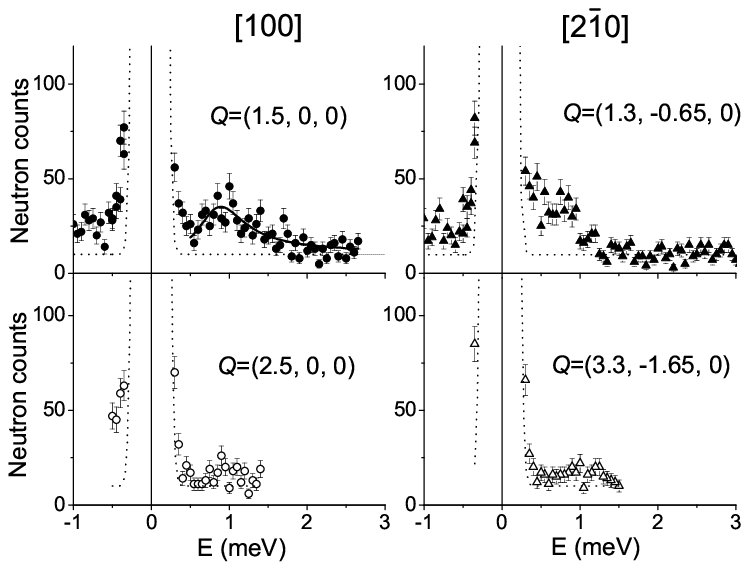}
\end{center}
\caption{Top : Intersection of the Brillouin zones with the (a*,
b*) scattering plane (dark gray lines). The white arrows indicate
the $Q$-scans performed in both the [100] and [2$\bar 1$0]
directions. Bottom : E-scans at 2 K for two different $Q$
positions in each direction (crosses in the top panel). The thin
lines are Gaussian fits of the incoherent peak. The thick line is
a lorentzian fit (see text).} \label{Escan}
\end{figure}

Scans in neutron energy transfer (E-scans) at different positions
along the [100] and [2$\bar 1$0] directions are shown in fig.
\ref{Escan}. At $Q$ = (1.5~0~0) (along [100]), a small signal is
visible around 0.9 meV. At $Q$ = (1.3~-0.65~0) (along [2$\bar
1$0]) a small signal is also present at a slightly smaller energy
and mixed with the tail of the incoherent peak. Along each
direction, the intensity of the signal decreases at larger $Q$
values, as expected for magnetic excitations, which are multiplied
by the square of the magnetic form factor \cite{footnote2}. Phonon
excitations would have shown a rise of intensity with $Q^2$. A
half width at half maximum (HWHM) $\Gamma$ = 0.47~meV is deduced
from the fit of the energy response at $Q$=(1.5~0~0), with a
lorentzian times the detailed balance factor, which corresponds to
a life time $\tau$ = 1.4~10$^{-12}$~s for the spin correlations.

\begin{figure}[t]
\begin{center}
\includegraphics[scale=1]{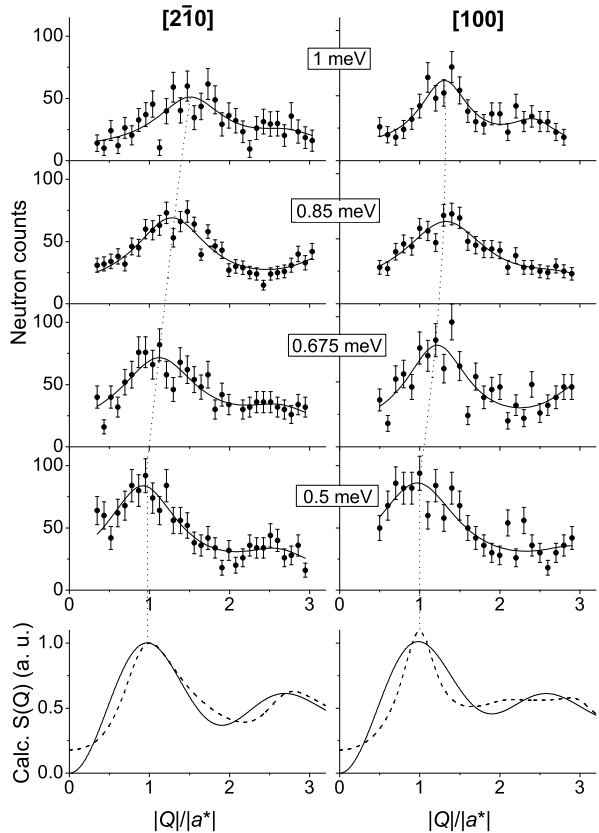}
\end{center}
\caption{$Q$-scans at 2 K at several energy transfers, along the
[100] and [2$\bar 1$0] directions. The dotted lines indicate the
position of the first maximum. The solid lines following the
measured $Q$ distributions (circles) are lorentzian fits (see
text). At bottom, static scattering, calculated using first
neighbors magnetic correlations only (solid line) and in a model
involving further neighbors correlations (dashed line), are
plotted for the same directions for comparison (see text).}
\label{Qscanall}
\end{figure}

Scans in neutron wave vector transfer ($Q$-scans), along the [100]
and [2$\bar 1$0 ]directions, at the 0.5, 0.675, 0.85 and 1~meV
energy transfers, show a weak, gradual and global evolution (cf.
fig. \ref{Qscanall}). The signal is minimum around $|Q|/|a^*|$ =
2, between a broad maximum and (in most scans) a second weaker
maximum. The first maximum is also observed in the intermediate
direction [4$\bar 1$0] at the same $Q$ value than in the other two
directions, confirming the peculiar ring shape of this signal
\cite{robert}. Its amplitude decreases and its position shifts
from $|Q|/|a^*|$ = 1 to 1.5 with increasing energy. This indicates
a significant dispersion with $Q$ of the magnetic scattering in
the 0.5-1~meV range. The position and amplitude of a second
maximum, when present, also vary with energy. The lines in fig. 2
show the results of fits to two lorentzians of same width,
multiplied by the square of the Nd$^{3+}$ magnetic form factor
\cite{brown}. The fitted HWHM $\kappa \approx$ 0.49 \AA$^{-1}$
corresponds to a correlation length $\lambda \approx$ 2 \AA,
smaller than the distance between the nearest neighbors Nd$^{3+}$
ions (4.2 \AA).

\begin{figure}[b!]
\begin{minipage}[t]{4cm}
\begin{center}
\includegraphics[angle=90,bb=100 22 350 258,scale=0.4]{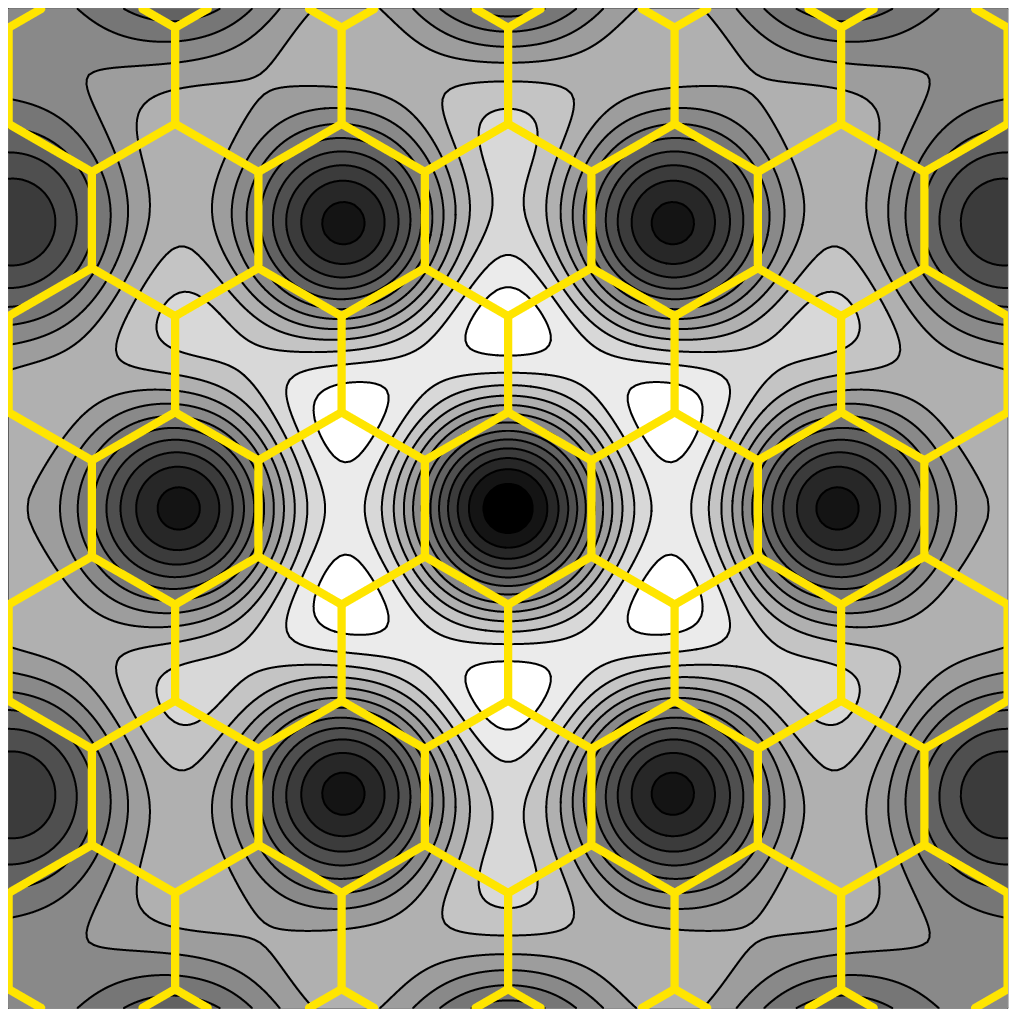}
\end{center}
\end{minipage}
\begin{minipage}[t]{4cm}
\begin{center}
\includegraphics[angle=90,bb=100 22 256 271,scale=0.4]{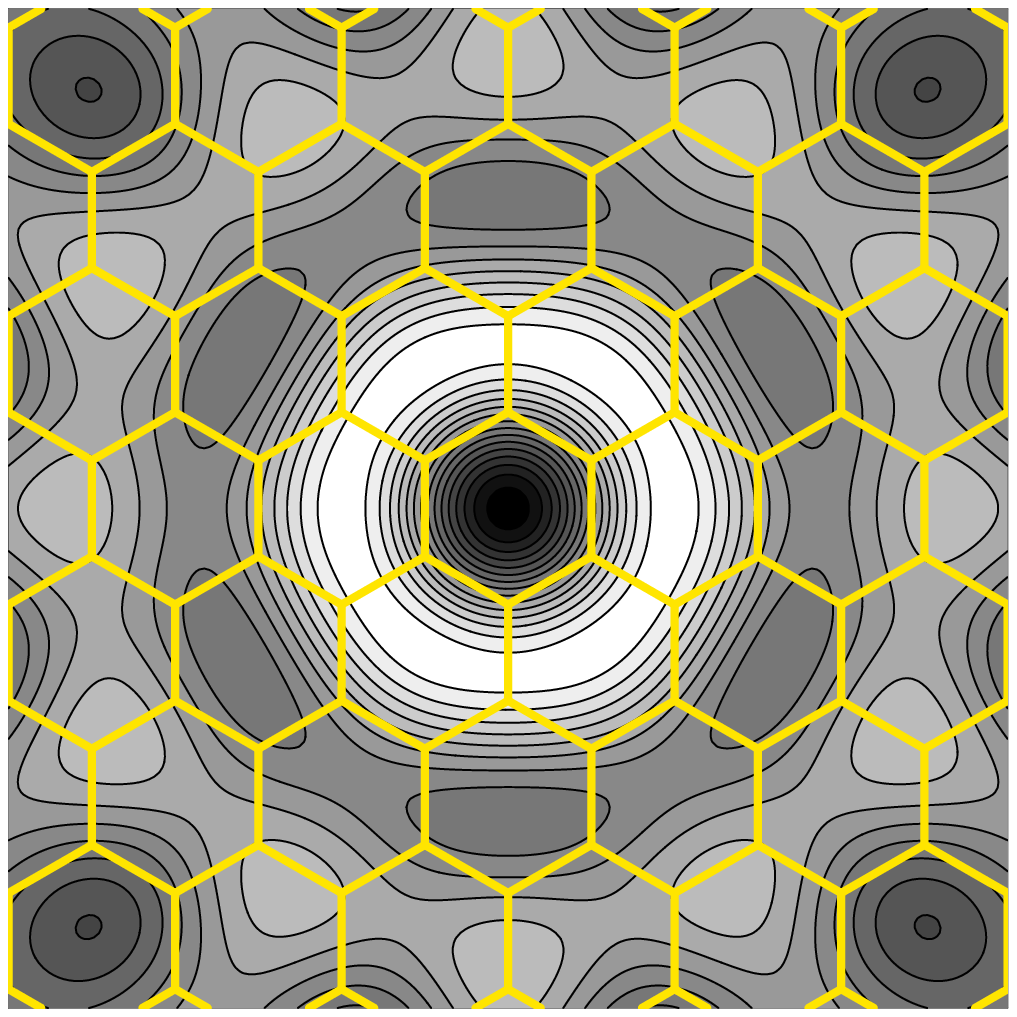}
\end{center}
\end{minipage}
\caption{Calculated static scattering map for the
antiferromagnetic kagom\'e lattice (left) and for the distorted
Nd-langasite kagom\'e lattice (right) using first neighbor spin
correlations (see text). The brightness increases with intensity.}
\label{map1}
\end{figure}

An interpretation of these results can be proposed from the
comparison of the measured magnetic signal with calculated neutron
static scattering intensities using different models of
antiferromagnetically interacting classical spins on the distorted
kagom\'e lattice formed by the Nd$^{3+}$ ions. In all cases, the
calculated Fourier-transformed spin correlation functions have
been multiplied by the square of the Nd$^{3+}$ magnetic form
factor \cite{brown}. The simplest model is obtained considering
correlations between first neighboring spins only and spin
configurations minimizing the energy on each triangle of the
structure. These are given as having the sum of three spins equals
to zero for Heisenberg or XY spins and two spins in one state and
the third spin in the opposite state for the Ising spins.
Surprisingly, the resulting $Q$ distribution of magnetic intensity
does not depend on the spin degrees of freedom. This validates the
interpretation of the experiment with this model albeit the low
temperature anisotropy of the compound is not quantitatively
determined at present. It is shown in fig. \ref{map1} where it is
compared to the same calculation on the non-distorted kagom\'e
lattice. The magnetic intensity distribution in the scattering
plane is more isotropic in the distorted case, consistently with
the experimental data. A second model, describing an algebraic
spin liquid at $T=0$, was derived using classical isotropic
infinite-component spin vectors following the method of ref.
\onlinecite{garanin}. The $Q$ distribution of magnetic intensity
was calculated at a finite temperature damping the dipolar-like
long-range spin-spin correlation by an exponentially decaying
term, hence smoothing the sharp features of the T = 0~K
calculation. The main features of the measured $Q$-distribution
are reproduced by both models : a first broad maximum localized
about $|Q|/|a^*|$ = 1 from the origin and a second one, less
well-defined, around $|Q|/|a^*|$ = 2.6. As expected from a static
calculation, the best agreement between the measured and
calculated $Q$-distributions is obtained at the lowest energy, 0.5
meV, at least concerning the position of the first maximum which
essentially reflects the first neighbors antiferromagnetic
correlations (cf. fig. \ref{Qscanall}).

The measured $Q$-distribution is consistent with a disorder phase
with specific magnetic correlations, the degeneracy of which would
result in apparent very short correlation lengths. Crudely
enlightened by the models, its main features are the broad ring
shape scattering about $|Q|/|a^*|$ = 1, the scattering empty
Brillouin zone around the origin, and the scattering minima in
Brillouin zones around certain reciprocal lattice points (cf. fig.
\ref{map1}). They can be indeed understood as expressing the
coexistence of many configurations at the same energy with
frustration constrained antiferromagnetic correlations. A similar
interpretation was invoked for the Y$_{0.97}$Sc$_{0.03}$Mn$_2$
spin-liquid, the short-range magnetic correlations of which would
originate from 4-site collective spin singlets associated with the
tetrahedral building unit of the pyrochlore lattice. Note also the
similarity of the $Q$-scans recorded in both Nd-langasite and
Y$_{0.97}$Sc$_{0.03}$Mn$_2$ \cite{ballou} which can be related to
the fact that the kagom\'e lattice is obtained through a cut
perpendicular to the cube diagonal of the pyrochlore lattice. A
basic difference is that no dispersion of the magnetic signal was
observed in Y$_{0.97}$Sc$_{0.03}$Mn$_2$, owing to a much shorter
lifetime of the spin correlations, $\approx$ 7 10$^{-14}$~s, so
that the modes either have no time to propagate or are drown
within the soft modes of the ground manifold.

Spin waves can propagate in locally ordered regions if the system
retains a sufficient temporal and spatial stiffness. In a
disordered medium, the spin waves lifetime is reduced, not only
through thermal fluctuations ($\tau_{th}$) \cite{McLean} but also
through the time scale on which the medium fluctuates
(autocorrelation time $\tau_a$) \cite{moessner98}, both of the
order of $\hbar / k_B T$. The mixing of the excitations built on a
degenerate ground state manifold further reduces their lifetime to
approximately $\hbar / \left[ J k_B T \right]^{1/2}$ (see
ref.~\onlinecite{moessner98}). But at the temperature we are
interested in, we have  $\tau_{th} \sim \tau_a \sim 3.8~10^{-12}$
s and a measured lifetime $\tau \sim 1.4~10^{-12}$ s. In this
regime, the reduction due to the degeneracy of the ground state is
small and the dynamics may still be seen as conventional e.g, the
spin wave spectrum should be built as a superposition of all
excitations developing over each magnetic configuration of the
ground state manifold. As a consequence, the density of quasi
static excitations should be much larger than the one at finite
energy because of the soft modes. Assuming a lorentzian shape with
a HWHM $\sim 0.2$~meV (from $\tau_a$) for their description, it is
reasonable to state that these quasi-static excitations also
contribute to the scattering function at finite energy transfer.
This would explain why the $Q$-dependence of the scattering is
well accounted for by the calculated static scattering
intensities, especially at the lowest energy transfer (0.5~meV).
Quasi static contributions from soft modes have a smaller weight
at higher energies and propagating modes become dominant, as
evidenced from the dispersion of the position of the first maximum
in the $Q$-distribution. Now, a spatial propagation is expected
for excited modes with wavelength smaller than the local order
correlation length. The one deduced from the $Q$-dependence of the
scattering ($\xi \sim 2$~\AA) seems too short to produce the
observed dispersion. This apparent contradiction can be lifted
considering that the correlation length associated to each
configuration of the ground state manifold would be large enough
to allow propagation of excited modes, whereas the measured one
reflects the distribution of close configurations. The conjugate
observation of a very broad signal associated to a dispersion of
its maximum could thus be a consequence of the high degeneracy of
magnetic states in this frustrated compound.

In conclusion, signatures of a cooperative paramagnet were
evidenced in Nd$_3$Ga$_5$SiO$_{14}$, in particular a peculiar $Q$
distribution of the magnetic correlations that would emerge from a
highly degenerate manifold of configurations. Although the unusual
dispersion of the magnetic correlations and the role of the
single-ion magnetocrystalline anisotropy have to be further
clarified, the compound provides with the first measurements of
the spatial and temporal magnetic correlations in a spin liquid
phase of a geometrically frustrated kagom\'e antiferromagnet.

\begin{acknowledgments}

We are grateful to A. Ibanez for his contribution during the
preparation of the Nd-langasite powder and to J. Balay and A.
Hadj-Azzem for their help in the synthesis of the Nd-langasite
single-crystal. We also thank M. Zhitomirsky for his critical
reading of the manuscript prior to publication.

\end{acknowledgments}


\begin{thebibliography}
\bibitem{footnote2} More rigorously, the magnetic intensities should be modulated by
the magnetic structure factor. The good agreement of the measured
$Q$-scans with the calculated magnetic structure factors is
therefore a further proof of the magnetic origin of this signal.
\end{thebibliography}
\end{document}